\documentclass[a4paper,twoside, fleqn]{article}
\newcommand{\affil}[1]{$^{\rm #1}$}
\usepackage[authoryear]{natbib}
\bibpunct{(}{)}{;}{a}{}{,}
\usepackage{graphicx}

\date{} 
\title{\large\bf\flushleft Recurrent $\sim$24 h Periods in {\it RXTE} ASM Data}
\author{\parbox{\textwidth}{\flushleft 
\vspace{-0.5cm}
{\it S. A. Farrell\affil{A,C}, P. M. O'Neill\affil{B}, and R. K. Sood\affil{A}}\\
\vspace{0.4cm}
{\small \affil{A}\,School of PEMS, UNSW@ADFA, Northcott Drive, Canberra, ACT 2600, Australia}\\
{\small \affil{B}\,Astrophysics Group, Imperial College London, Blackett Laboratory, Prince Consort Rd, 
London SW7 2AZ, United Kingdom}\\
{\small \affil{C}\,Email: s.farrell@adfa.edu.au}}}
\begin{document}
\twocolumn[
\begin{changemargin}{.8cm}{.5cm}
\begin{minipage}{.9\textwidth}
\vspace{-1cm}
\maketitle
%
%
\small{\bf Abstract:}
Analysis of data from the {\it Rossi X-ray Timing Explorer} satellite's All Sky Monitor instrument for several X-ray binary sources has identified a recurrent $\sim$24 h period. This period is sometimes highly significant, giving rise to the possibility of it being identified as an orbital or super-orbital period. Further analysis has revealed the same period in a number of other X-ray sources. As a result this period has been discounted as spurious, described variously as arising from daily variations in background levels, the scheduling of ASM observations and beating between the sampling period and long-term secular trends in the light curves. We present here an analysis of the spurious periods and show that the dominant mechanism is in fact spectral leakage of low-frequency power present in the light curves.


\medskip
\medskip
\end{minipage}
\end{changemargin}
]
\small

\section{Introduction}
The {\it Rossi X-ray Timing Explorer} ({\it RXTE}) spacecraft was launched on 1995 December 30 into a circular low-inclination ($\sim$23$^\circ$), Low Earth Orbit ($\sim$580 km), with an orbital period of $\sim$96 min (Jahoda et al. 1996).  It continues to operate almost flawlessly up to the present time. The payload consists of two pointed instruments --– the Proportional Counter Array (PCA) and the High Energy X-ray Timing Experiment (HEXTE); and one scanning instrument –-- the All Sky Monitor (ASM).

Analysis of ASM data using a variety of techniques has identified a number of spurious periods. Power at the $\sim$96 min spacecraft orbital period is evident in the power spectra of most sources, with other spurious periods reported at $\sim$180 d and $\sim$365 d (Benlloch 2004). Additional periods at $\sim$50 d and $\sim$11 yr might be expected due to the precession of the spacecraft orbit (Corbet 2003), and from modulations due to the solar cycle, although the ASM data sets are not long enough to see an $\sim$11 yr periodic variation as anything other than a long-term secular trend. A highly significant period at $\sim$24 h has also been reported for a number of sources and has been variously attributed to daily variations in the X-ray background (Corbet et al. 1999), beating between the sampling period of the satellite and a long-term secular trend in the light curve (Benlloch 2004), and to the scheduling cycle of the ASM (Zdziarski et al. 2004). At least two X-ray binary systems are suspected to have orbital periods around this value: GX 339-4 ($\sim$0.5 –- 1.7 d; Buxton \& Vennes 2003; Hynes et al. 2003) and CAL83 (1.0436 d; Smale et al. 1988). While variations at the orbital period in X-rays do not appear to be evident in {\it RXTE} ASM data for either source, it is entirely possible that other sources exist which exhibit real variations around this period. It is thus important that the underlying mechanism be determined so that spurious periods can be confidently ruled out.

\section{The All Sky Monitor}
The ASM consists of three wide-angle Scanning Shadow Cameras (SSCs) covering a nominal 1.5 -- 12.0 keV energy range with three energy channels (1.5 -- 3.0, 3.0 -- 5.0, and 5.0 -- 12.0 keV; A. M. Levine 2004, private communication). The ASM assembly scans $\sim$80$\%$ of the sky every orbit, "dwelling" on a target source for $\sim$90 s each observation (Bradt, Rothschild \& Swank 1993). Each camera consists of a position sensitive xenon proportional counter covered by an 8 $\mu$m thick aluminized plastic thermal shield and a 50 $\mu$m thick beryllium window. In addition, the proportional counters in SSCs 2 and 3 have a $\sim$2 $\mu$m thick coating of polyimide on the interior of the beryllium foil as a protection against leaks (Levine et al. 1996). The field of view of 6$^\circ$ $\times$ 90$^\circ$ FWHM is limited by a thin aluminium coded mask collimator. The current target catalogue contains over 300 X-ray sources, each observed up to 15 times per day by one or more of the SSCs. As a result, the ASM light curves can contain a number of data points within any given orbit, as the source moves in and out of the fields of view of the three SSCs.

The ASM collects data with a duty cycle of $\sim$40$\%$ which, when combined with spacecraft manoeuvres, results in a highly stochastic pattern of sky coverage (Levine et al. 1996).  The resulting observing pattern for an individual source is thus uneven, with the orbital location of {\it RXTE} while observing a specific source seeming to change randomly over time. The pattern of the shadow cameras coded mask along with a model of the diffuse X-ray background is taken into account, with the source intensities obtained from the solution of a linear least-squares fit. Corrected intensities from individual dwells with high values of reduced $\chi^2$, short exposure times due to entry into high-background regions, from sources close to or behind the Earth's limb, from sources close to the edge of the field of view, and from fields that are exceptionally crowded are rejected (Levine et al. 1996). The resulting background subtraction is not perfect, with numerous negative count rates appearing in each light curve. The corrected intensities from individual observations (or dwells) are compiled and made available to the public domain on NASA's High Energy Astrophysics Science Archive Research Centre (HEASARC) website\footnote{http://heasarc.gsfc.nasa.gov/docs/xte/xtegof.html}. The ASM has been operating on a nearly continuous schedule, turning off only during passage through high-background regions such as the South Atlantic Anomaly (SAA) and whenever the Sun is in the field of view (Levine et al. 1996).

Other systematic errors that may affect some sources include effects upon the intensity when another bright source is in the field of view, incompleteness of the active catalogue of X-ray sources, and low energy absorption by the window and thermal shield increasing with elevation of a source in the field of view. In addition, a slow drift in the gain in SSCs 1 and 3 has been observed over time, along with a change in the channel definitions used to define the three energy bands which occurred at MJD 51548 (Corbet 2003). SSC 3 has sustained the most degradation over time, while the gain in SSC 2 has remained relatively stable.

\section{The $\sim$24 h Period}
\subsection{Data Analysis and Reduction}
{\it RXTE} ASM background-subtracted single dwell data were used for all our analyses. As the observed 24 h periods are not believed to originate from the distant source, a barycentric correction was not performed. Data from SSC 3 were rejected due to the degradation of the instrument, and linear trends were fitted to the light curves before and after the gain adjustment in MJD 51548. These linear trends were then subtracted to remove the resulting spurious low-frequency power. All individual dwells with an error greater than 1$\sigma$ from the mean error in the light curve were removed, along with any dwells with anomalously low count rates. The mean count rate was then subtracted from the light curve to reduce noise in the power spectra.

The primary analysis technique employed on the reduced light curves was the Lomb-Scargle periodogram (Lomb 1975; Scargle 1982), although epoch folding and epoch folding search techniques using $\chi^2$ statistics were also employed to test that the problem was not a result of the analysis technique alone. The Lomb-Scargle power spectrum is particularly useful in the identification of periodic signals buried in noisy unevenly spaced time series, while the epoch folding search technique is best suited for the identification of non-sinusoidal signals. Epoch folding of a light curve at a suspect period can then determine the shape and parameters of a periodic signal. All three of these techniques detected the presence of $\sim$24 h modulation in the suspect light curves, indicating that the spurious period is not an artefact of a particular analysis technique. For the periodogram analyses we employed the \textsc{fasper} subroutine (Press \& Rybicki 1989), while the epoch folding and epoch folding search techniques made use of the \textsc{ftools} analysis software available through the HEASARC website. It should be noted that the False Alarm Probability (FAP) of a given peak in a Lomb-Scargle power spectrum depends on the sampling pattern of the data (Horne \& Baliunus 1986). Therefore, we used Monte Carlo simulations (e.g. Kong, Charles \& Kuulkers 1998) to determine the 99$\%$ white noise significance levels for all power spectral analyses.

\subsection{Characterisation of the $\sim$24 h Period}
During the course of analysis of ASM light curves, we first identified a spurious period around 24 h for the Low Mass X-ray Binary black hole candidate GX 339-4. This happened while searching for its previously reported orbital period of between $\sim$0.5 -- 1.7 d. Highly significant power was observed in the Lomb-Scargle power spectra around 24 h, which we provisionally associated with the orbital period. However, further analysis of ASM data revealed four other sources within 5$^\circ$ of GX 339-4 that also showed power at $\sim$24 h. Benlloch (2004) had also noted the 24 h period in the GX 339-4 light curve, attributing it to beating between the sampling period of the spacecraft and a secular trend in the light curve. Zdziarski et al. (2004) on the other hand pointed to the scheduling cycle of the {\it RXTE} observations as the cause of the presence of this period.

Furthermore, Corbet, Finley \& Peele (1999) had identified a similar 25.65 h period in the High Mass X-ray binary pulsar 2S 0114+650, and had explained it as being related to daily variations in background levels. They had also noted that variability at similar periods had been seen in at least some other sources observed with the ASM. The observed variability in 2S 0114+650 was attributed to passage of the spacecraft through the SAA (R. H. D. Corbet 2004, private communication). While this 25.65 h period is clearly present in the $\sim$2 yr of ASM data that Corbet, Finley \& Peele utilised, it is not significantly present in our 8.5 yr data set (see Farrell, O'Neill \& Sood   2005). Subsequently, many other sources have been revealed that show power at $\sim$24 h.

Perhaps the most puzzling aspect of the 24 h period is that it is not present in data from all sources --– it seems to appear only in conjunction with significant low-frequency power, although there are cases where sources with confirmed low-frequency modulation do not show significant power at or around the suspect period. The sources with observed 24 h periods are not all grouped together in the sky but tend to vary in their celestial coordinates. Furthermore, the value of the period is not constant between sources, and is sometimes seen to "split" forming a double peak either side of 24 h. The power of the period varies between sources and even within a single light curve, sometimes disappearing completely in certain sections of a data set.

\subsection{Sources with $\sim$24 h Periods}
\begin{figure}[htb]
\begin{center}
\includegraphics[scale=0.3, angle=270]{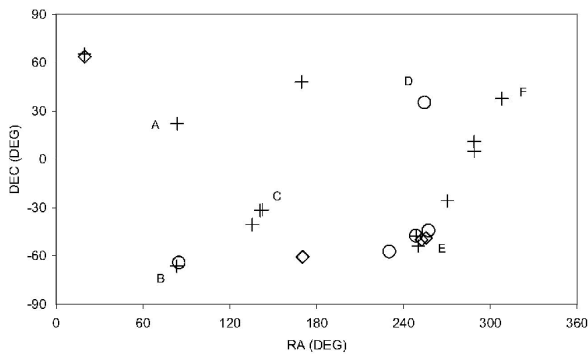}
\caption{Location of 22 sources identified as having $\sim$24 h periods, ranked in terms of strong (diamonds), medium (circles) and weak (crosses) signal/noise ratio of the 24 h peaks. Sources used for our analyses were: (A) Crab pulsar; (B) LMC X-4; (C) Vel X-1; (D) Her X-1; (E) GX 339-4; and (F) EXO 2030+375}\label{fig1}
\end{center}
\end{figure}

\begin{figure}[h!tb]
\begin{center}
\includegraphics[scale=0.325, angle=0]{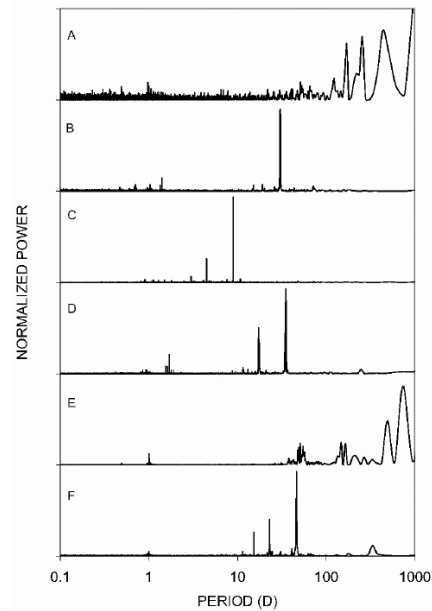}
\caption{Power spectra of the six test sources using all available {\it RXTE} ASM data. Each source is referenced using the same letter denoting its location in Figure~\ref{fig1}}\label{fig2}
\end{center}
\end{figure}

\begin{figure}[h!tb]
\begin{center}
\includegraphics[scale=0.325, angle=0]{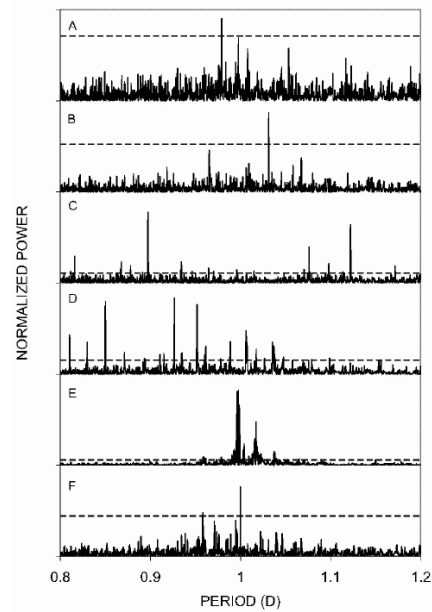}
\caption{Power spectra of the six test sources using all available {\it RXTE} ASM data showing the 24 h peaks. The 99$\%$ significance levels are indicated by the dashed lines}\label{fig3}
\end{center}
\end{figure}

We have identified 22 X-ray sources with power at $\sim$24 h in their power spectra. Figure~\ref{fig1} shows the location in the sky of these sources, grouped into sources with strong, medium, and weak power at 24 h. From these we selected six well studied sources randomly located in the sky for our analyses, each marked on Figure~\ref{fig1} and referred to in all subsequent figures with the corresponding letters: Crab pulsar (A), LMC X-4 (B), Vel X-1 (C), Her X-1 (D), GX 339-4 (E), and EXO 2030+375 (F) . Each of these selected sources showed evidence of the 24 h period with $>$ 99$\%$ significance in all available ASM data ($\sim$8 yr of data), and in a 400 d time period spanning MJD 52385 –- 52785. Figure~\ref{fig2} shows the power spectra of all available ASM data for each of these sources, covering 0.1 -– 1 000 d. While the low-frequency power present in the Crab power spectrum is unexpected, the power spectra of the other five sources matches well with previously published data, indicating that our analysis techniques are sound. The peaks present at periods $>$ 150 d in the Crab power spectrum are indicative of long-term variations in the light curve, possibly due to background variations arising from the low declination of the source and/or the solar cycle. Figure~\ref{fig3} is a higher resolution look at the 24 h peaks in the same power spectra covering 0.8 -– 1.2 d, with the 99$\%$ significance levels indicated.

\section{Daily Variations in \\Background Levels}
\subsection{SAA Passage}
The area of low magnetic rigidity known as the South Atlantic Anomaly has long been known to cause serious problems for spacecraft. The SAA is believed to be a result of the magnetic dipole axis being offset from the centre of the Earth, resulting in a comparative weakening of the geo-magnetic field off the coast of Brazil. An increase in the radiation levels at Low Earth Orbit altitudes can lead to single event upsets, surface charging, and component failures among other problems. As a result, most spacecraft routinely switch off critical subsystems during passage over the SAA in order to minimize disruption and damage.

The high energy of the charged particles in this region is often within the range of interest of satellite based X-ray and gamma-ray telescope instruments. The very high fluxes of these particles ($>$ 1 000 protons cm$^{-2}$ s$^{-1}$ above 30 MeV) result in the activation of isotopes within the spacecraft material during passage (Shaw et al. 2003). These isotopes have varying decay half-lives, sometimes lasting until well after the spacecraft has left the SAA region. During this decay time-scale the background radiation levels are increased, resulting in a higher count rate from instruments such as the proportional counters used in the ASM and the PCA. Observations made within the time frame of activation decay inherently introduce periodicities into the light curves such as the spacecraft orbital period and the rotational period of the Earth. {\it RXTE} for instance regularly passes through the SAA approximately 10 out of every 15 orbits, as the Earth rotates slowly under the orbital path (E. Smith 2004, private communication). Background levels due to activation will gradually increase with each orbit until the SAA moves out from underneath the orbital path. The background levels then slowly decrease for the 5 orbits until re-entry, hence introducing an approximate 24 h periodicity into the data. Another periodicity at $\sim$50 d may be introduced in a similar fashion as the orbit of {\it RXTE} gradually precesses at this period. After 50 d the time-of-day that SAA passages occur will have moved earlier through one whole day, resulting in 51 cycles of SAA passage within the 50 d (E. Smith 2004, private communication).

The differing decay half-lives of radioactive isotopes contribute separate components to the background levels of both the PCA and ASM instruments. The two most significant components in the PCA background have $\sim$25 and $\sim$240 min decay time-scales, although the specific contributing species have not been identified (C. Markwardt 2004, private communication). The instruments that make up the PCA consist of five xenon/methane proportional counters with a 90/10 mixture ratio and a propane veto level (Jahoda et al. 1996). Each ASM SSC consists of a xenon/CO$_2$ proportional counter with a 95/5 mixture ratio (Levine et al. 1996). As both instruments are xenon proportional counters, it is fair to assume they will experience similar activation decay time-scales. 

The main material components for both instruments are xenon and aluminium.  The background introduced into the detectors as a result of the passage through the SAA will be a function of the relevant excitation cross-sections for these elements, the particle flux in the SAA, and the half-lives of the radio-isotopes consequently produced.  Aluminium does not have radio-isotopes that have  characteristic activation-decay time-scales of $\sim$25 or $\sim$240 min.  120Xe, 121Xe and 123Xe have half-lives of 40, 39, and 120 min respectively, but it is highly improbable that these isotopes will be produced in the expected scheme of excitation, as the most abundant naturally occuring isotopes of xenon have atomic mass numbers of 129 or greater.

\begin{figure}[h!tb]
\begin{center}
\includegraphics[scale=0.325, angle=0]{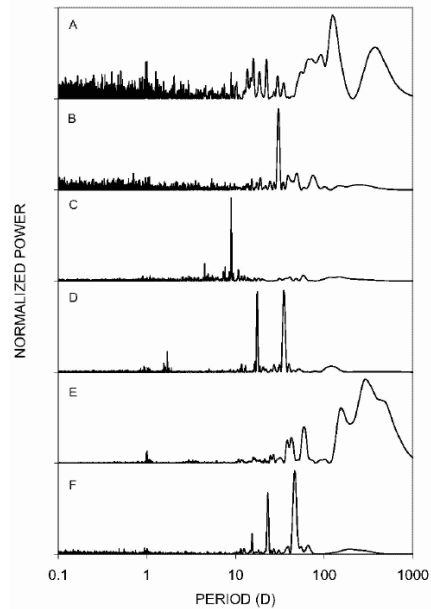}
\caption{Power Spectra of the 400 d light curves spanning MJD 52385 -– 52785 for the six test sources}\label{fig4}
\end{center}
\end{figure}

In order to investigate SAA passage as the root mechanism for the observed 24 h periods, we first modelled the location of the SAA with respect to the spacecraft over time. To do this we made use of the {\it RXTE} orbital ephemeris data file available via the HEASARC website. Due to the high time resolution of these data (the orbit file contains the location and velocity of the spacecraft every 60 s since launch) and the size of the data file, we selected a 400 d section spanning MJD 52385 -– 52785 for our analysis. Using the {\it RXTE} housekeeping data from PCA observations of 2S 0114+650, we were able to determine the location of the SAA relative to the spacecraft at one particular instance in time. The housekeeping files give the spacecraft location in Earth latitude and longitude against mission time, allowing us to calculate the RA and dec of the SAA for a given time stamp. Taking the coordinates of the SAA maximum as 25$^\circ$ S, 45$^\circ$ W (Plett 1978) we were able to model its movement by keeping the dec constant and stepping the RA parameter with the 23.9345 h rotational period of the Earth. In order to validate our model we referred to PCA observations of the Crab pulsar, Cir X-1 and 2S 0114+650 listing the time since SAA passage for each individual time increment. We compared the location of the SAA obtained from our model with that of the spacecraft during small values of the time since SAA parameter, and found our model to be a good fit for all three sources over the 8.5 yr mission life of {\it RXTE}.

\begin{figure}[h!t]
\begin{center}
\includegraphics[scale=0.325, angle=0]{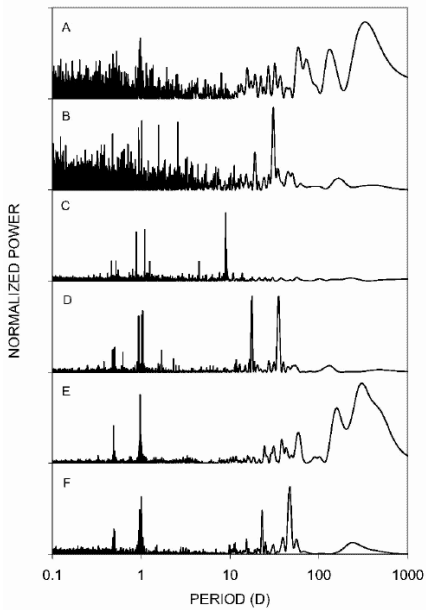}
\caption{Power spectra of the 400 d light curves after SAA affected observations were filtered out}\label{fig5}
\end{center}
\end{figure}

We generated power spectra for the six test sources using the Lomb-Scargle periodogram technique for the reduced ASM data from SSCs 1 \& 2, between MJD 52385 -– 52785 (Figure~\ref{fig4}). Applying our model of the SAA coordinates, we generated a list of times during which {\it RXTE} was within a conservative region of $\pm$ 45$^\circ$ RA and $\pm$ 25$^\circ$ dec centred on the location of the SAA maximum. A new 'time since SAA passage' parameter was then calculated for each individual dwell time stamp for each of the six test sources. These parameters were used to filter out observations that were made within a given time ($\Delta$T) of the spacecraft leaving the SAA affected region. Filtering out all observations within an initial value of 30 min for $\Delta$T and recalculating the power spectra, we found that the 24 h period was not significantly reduced for any of the sources. Repeated analysis with a $\Delta$T of 120 min should have seen significant reduction in the power of the 24 h peak, due to the removal of all observations within one orbit of SAA passage. Filtering with a $\Delta$T of 120 min would also remove all observations possibly affected by the decay of any xenon isotopes activated by SAA passage. Our results show that this is not the case --– not only were the peaks still present they had significantly increased in power. Figures~\ref{fig4} \& \ref{fig5} show the power spectra for each of the test sources before and after SAA filtering. While it is possible that the decay of an isotope with a longer half-life (such as the 240 min component mentioned previously) could be causing the 24 h period, this is unlikely due to an additional factor. Simulated power spectra of reported super-orbital periods present in a number of the affected sources (e.g. Her X-1 \& GX 339-4) show power around 24 h, indicating that the spurious period is most likely not an artefact of the instrument count rates but instead related to the sampling period. 

\subsection{An `Anti-Anomaly'?}
\begin{figure}[ht!]
\begin{center}
\includegraphics[scale=0.325, angle=0]{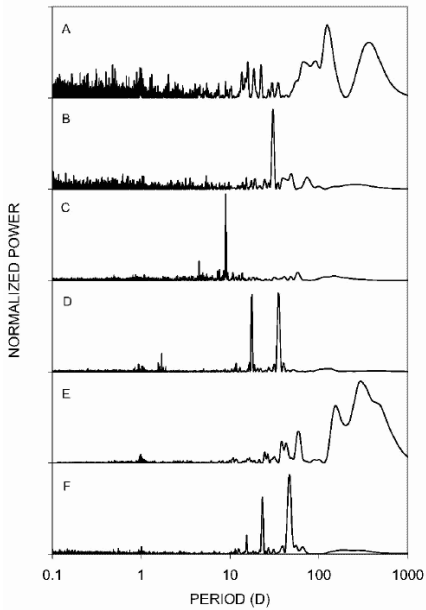}
\caption{Power Spectra of the 400 d light curves after AA affected observations were filtered out}\label{fig6}
\end{center}
\end{figure}

While the offset of the geo-magnetic dipole results in an area of low magnetic rigidity at the SAA, an area of high magnetic rigidity must exist on the opposing hemisphere, which we will refer to as the Anti-Anomaly (AA). Passage through the AA would result in lower levels of background radiation due to increased shielding from the geo-magnetic field. This could result in periodic dips in the background levels, possibly manifesting in the power spectra as peaks similar to those described previously for the SAA passage. Spacecraft instruments and subsystems are not commonly shut down during AA passage so a number of observations are bound to occur over the AA each orbit. The issue of isotope activation is not present in this case due to the reduction in the flux of high energy charged particles. Taking the location of the AA as 10$^\circ$ N, 100$^\circ$ E (Plett 1978), we modelled the movement of the AA in a similar fashion to the method previously described for the SAA. Using the same six test sources, we filtered out all observations made within a region of $\pm$ 45$^\circ$ RA and $\pm$ 25$^\circ$ dec centred on the AA location. Figure~\ref{fig6} shows the power spectra for each of the test sources after AA filtering. The power spectra shown in Figure~\ref{fig6} are practically indistinguishable from the unfiltered power spectra shown in Figure~\ref{fig4}, indicating that AA filtering did not significantly reduce the power of the 24 h peaks and thus is not the dominant mechanism.

\section{Spectral Leakage}
\subsection{ASM Sampling Rate}
The ASM observing schedule is determined by a software package that takes into account a number of variables, generating observing plans for one day time intervals. These variables include the pointing program of the spacecraft, the spacecraft orbital ephemeris, the location of the Sun and the apparent direction of the Earth, the limits of the ASM drive assembly rotation range, the slew rate of the drive assembly, and the locations of high background regions. The daily observing plans are highly complex, such that the times of observation of a given source tend to look random (A. M. Levine 2004, private communication). The sampling rate for a given source is partially uneven, with blocks of unevenly spaced observations reoccurring on a quasi-periodic basis. It is possible and even probable that this grouping has a period of around 24 h.

In order to test whether the sampling rate of ASM observations is the cause of the spurious periods, we simulated light curves with the same sampling rate of the affected sources but with random intensity values. The resultant Lomb-Scargle power spectra for each of the sources showed no significant power at any period, indicating that the sampling rate alone cannot be the cause of spurious periods at 24 h or any other period. When the random intensity values were replaced with a 30 d period sine wave, power at 30 d and $\sim$1 d emerged in all the power spectra. Closer examination of the $\sim$1 d peaks resolved them into two closely spaced peaks. Repeating the simulations with different input periods found that the spacing between the 1 d peaks was inversely proportional to the value of the input period, so that as the input period was reduced the peaks moved apart, and as it was increased the peaks converged. These results strongly imply that the spurious 24 h periods are not related to variations in background levels, but instead result from an interaction between the sampling rate and real modulations present in the light curves.

\subsection{Spectral Leakage Effects}
Spectral leakage is the transferral of power from a certain frequency to either higher or lower frequencies in a time series. Leakage to nearby frequencies is due to the finite length of the time series, while leakage to distant frequencies is a result of the finite size of the interval between samples (Scargle 1982). Unevenly sampling a data set tends to reduce and often eliminate the effect of spectral leakage. While astronomical data by the nature of the observations are typically unevenly sampled, there is sometimes an enforced regularity which results in semi-regular spacing of data points (Scargle 1982). Events that can enforce regularity in the case of the {\it RXTE} ASM include Earth occultation of the source, regular shut-down during passage through high background regions (e.g. the SAA), and the orbit of the Earth around the Sun. This enforced regularity can quite often lead to spectral leakage of a signal to other frequencies, and in fact explains why filtering of all SAA affected dwells increased the power of the 24 h peaks. Removal of all observations within 120 min of an SAA passage inherently increases the enforced $\sim$24 h regularity, increasing the leakage of power.

Benlloch (2004) found that leakage from low to high frequencies does occur in ASM light curves owing to the sampling of the data. While noting the general cause of the 24 h period in GX 339-4, the pseudowindow function of the data was not examined and the effect was not explored in-depth by this author. As such the appearance of only a single peak around 24 h was predicted, as opposed to the double peak structure that we have observed.

\begin{figure}[htb]
\begin{center}
\includegraphics[scale=0.3, angle=270]{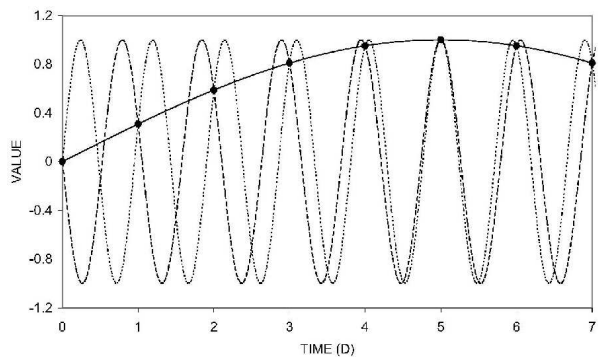}
\caption{Three sine waves with frequencies of 0.05 d$^{-1}$ (solid line), (1 + 0.05) d$^{-1}$ (dotted line) and (1 $-$ 0.05) d$^{-1}$ (dashed line), sampled at 1 d time intervals. All three sine waves can be reproduced by the observations indicated by the solid dots, demonstrating the effect of spectral leakage}\label{fig7}
\end{center}
\end{figure}

Spectral leakage can be illustrated with the following simple example, where the data contain a signal with a period of 20 d (frequency = 0.05 cycles d$^{-1}$). If we sample this signal regularly once-per day (i.e. $dt$ = 1 d), we will obtain the set of measurements indicated by the data points (dots) in Figure~\ref{fig7}. These measurements could also be produced with (at least) two other sine curves, with frequencies of (1 + 0.05) and (1 $-$ 0.05) cycles d$^{-1}$. So, if we were to calculate the periodogram we would get spurious peaks at $1/dt$ $\pm$ 0.05 cycles d$^{-1}$. In the evenly sampled case one would only ever extend the power spectrum up to the Nyquist frequency of 0.5 cycles d$^{-1}$ (i.e. 0.5$\times$($1/dt$) cycles d$^{-1}$), so this spectral leakage would never actually be observed.

\begin{figure}[htb]
\begin{center}
\includegraphics[scale=0.3, angle=270]{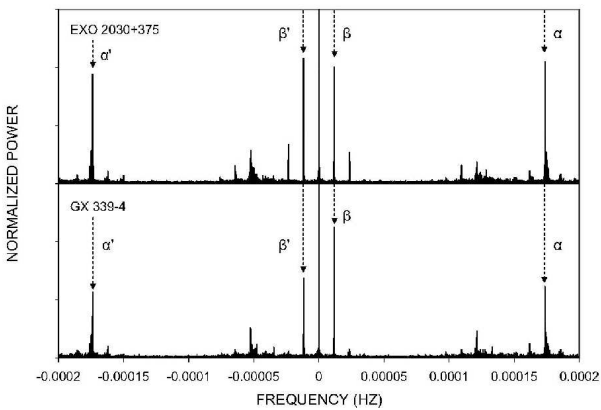}
\caption{Pseudowindows generated from the data on GX 339-4 (lower panel) and EXO 2030+375 (upper panel). Power from the {\it RXTE} orbital period of 96 min is marked by $\alpha$ (positive) and $\alpha$$^\prime$ (negative). The peaks due to the enforced $\sim$24 hr effective sampling rate are marked by $\beta$ (positive) and $\beta$$^\prime$ 
(negative)}\label{fig8}
\end{center}
\end{figure}

The 'pseudowindow' can be used to determine the response of a given data analysis system to a single frequency sine wave (Scargle 1982). All spectral leakage effects present in a given data set will manifest directly in the pseudowindow, indicating where any enforced quasi-periodicities lie in the sampling rate. Pseudowindows can be generated by simulating a single frequency sine wave using the same sampling rate as the source of interest. The frequency of the sine wave should be high enough so that spurious peaks from negative frequencies do not overlap with those from positive frequencies. A frequency of 2$\times$10$^{-4}$ Hz should be sufficient for most {\it RXTE} ASM data sets. The resulting Lomb-Scargle power spectrum of this simulated data set is the pseudowindow, and indicates where power will shift to due to spectral leakage. By translating the simulated power spectrum along the frequency axis so that the peak with the highest power (i.e. 2$\times$10$^{-4}$ Hz) lies at zero frequency, we can directly determine the effective sampling rates that are causing the observed spectral leakage. This observed power spectrum is thus the convolution between the true power spectrum and the pseudowindow. Figure~\ref{fig8} shows pseudowindows generated from the data on GX 339-4 and EXO 2030+375. Note that the power spectra are not completely symmetrical around zero frequency. After the zero frequency, the strongest power lies at 1.1604$\times$10$^{-5}$ Hz ($\sim$0.997 d) for GX 339-4 and -1.18275$\times$10$^{-5}$ Hz ($\sim$0.979 d) for EXO 2030+375, which are attributed to enforced $\sim$24 hr periodicities in the light curves (see Table 1 for the a list of the enforced $\sim$24 hr periods for all six test sources). The $\sim$96 minute orbital period of {\it RXTE} is also manifest in both power spectra at $\sim$$\pm$1.735$\times$10$^{-4}$ Hz.

\begin{figure}[htb]
\begin{center}
\includegraphics[scale=0.3, angle=270]{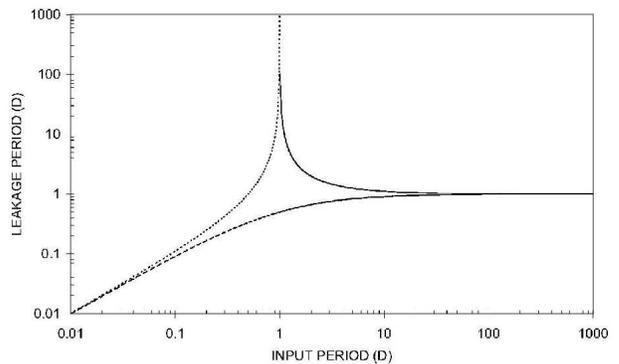}
\caption{The periods at which we expect to see spurious peaks, as a function of the input period, obtained from Equation (5) with a sampling period of 1 d}\label{fig9}\end{center}
\end{figure}

Pseudowindows are useful because they indicate where the spurious power will be seen for a certain sampling pattern. In the case of the {\it RXTE} ASM data, we will see spurious peaks at the input frequency ($F_{\mbox{in}}$) $\pm$ the `effective' sampling frequency ($1/dt$). So the spurious peaks ($F_{\mbox{s1}}$, $F_{\mbox{s2}}$) will be seen at:
\begin{eqnarray}
F_{\mbox{s1}} & = & F_{\mbox{in}} - (1/dt) \label{eq1}\\
F_{\mbox{s2}} & = & F_{\mbox{in}} + (1/dt) \label{eq2}
\end{eqnarray} 
For the case where the input frequency is less than the effective sampling frequency $F_{\mbox{s1}}$ is negative and so does not explain the appearance of two peaks within our frequency range, leading to the conclusion that we would only see $F_{\mbox{s2}}$. Taking into consideration that $F_{\mbox{in}}$ also produces power at $-F_{\mbox{in}}$, there are two more spurious peaks to consider:
\begin{eqnarray}
F_{\mbox{s3}} & = & -F_{\mbox{in}} - (1/dt) \label{eq3}\\
F_{\mbox{s4}} & = & -F_{\mbox{in}} + (1/dt) \label{eq4}
\end{eqnarray} 
As $F_{\mbox{s3}}$ will always be negative and thus not present in our frequency range, Equation (3) can be disregarded. That leaves three possible frequencies ($F_{\mbox{s1}}$, $F_{\mbox{s2}}$ or $F_{\mbox{s4}}$) that may be present. For the case where the input frequency is less than (1/$dt$), $F_{\mbox{s2}}$ and $F_{\mbox{s4}}$ will be present in the power spectrum. As $F_{\mbox{in}}$ approaches (1/$dt$), the two spurious peaks will move apart. As $F_{\mbox{in}}$ approaches zero, the two peaks will move closer together, converging at (1/$dt$). For the case where the input frequency is greater than (1/$dt$), $F_{\mbox{s1}}$ and $F_{\mbox{s2}}$ will be present in the power spectrum. We compute the Lomb-Scargle periodogram for positive frequencies only so we can summarise Equations (1) to (4) in a single equation:
\begin{eqnarray}
F_{\mbox{s}} & = & |F_{\mbox{in}} \pm (1/dt)| \label{eq5}
\end{eqnarray} 
Figure~\ref{fig9} shows the periods at which we expect to see spurious peaks, as a function of the input period.

\subsection{Results}
The effective sampling frequency 1/$dt$ can be determined directly from a pseudowindow generated from the data on each object. This value of 1/$dt$ can then be used in conjunction with Equation (5) to determine the frequency of the spurious peaks at $\sim$24 h in a given power spectrum with known period(s). Using this method, we can test whether the observed spurious 24 h periods are in fact a product of spectral leakage as described above. 

\begin{figure}[h!tb]
\begin{center}
\includegraphics[scale=0.325, angle=0]{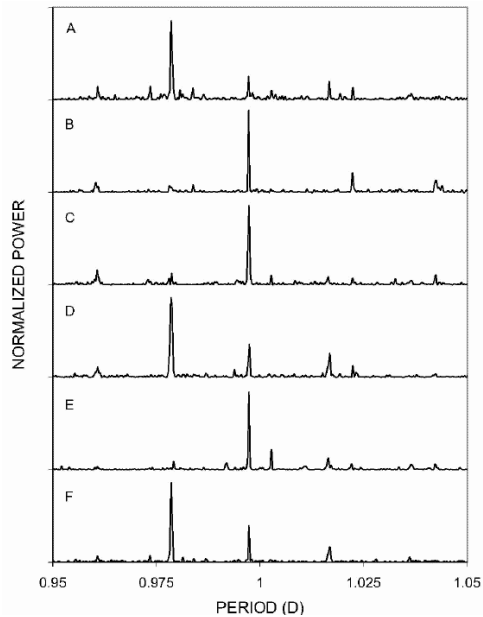}
\caption{Pseudowindows for the six test sources showing the $\sim$1 d effective sampling rate {\it dt}. The apparent similarity of the pseudowindows for sources A/D/F and B/C/E appear to be a result of the sign of the declination of the sources (i.e. northern or southern hemisphere)}\label{fig10}
\end{center}
\end{figure}

\begin{figure}[h!tb]
\begin{center}
\includegraphics[scale=0.325, angle=0]{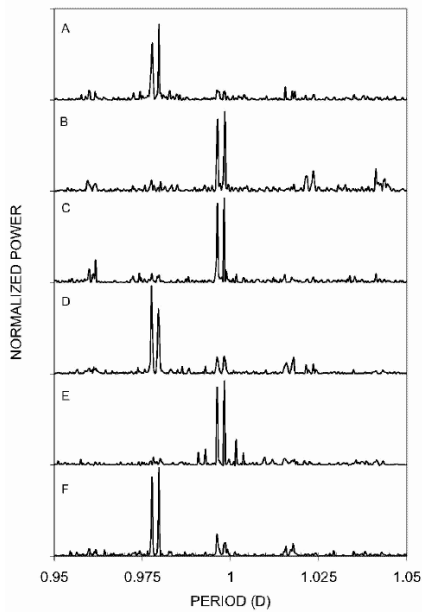}
\caption{Power spectra of simulated 1 000 d sine waves with the same sampling as the six test sources. The double peaks are located either side of the effective sampling period of each light curve as shown in Figure~\ref{fig10}}\label{fig11}
\end{center}
\end{figure}

To test this hypothesis, we simulated a 2$\times$10$^{-4}$ Hz sine wave sampled at the same rate as each of our test sources, in order to determine the effective sampling rate $1/dt$. Figure~\ref{fig10} shows pseudowindows between 0.95 -– 1.05 d for all six test sources, while the effective sampling periods are listed in Table~\ref{table1} with the uncertainties determined by taking the FWHM of the $\sim$1 d peaks. To compare these sampling rates with the actual spectral leakage peaks, we simulated 1 000 d sine waves with the same sampling rates. The resulting power spectra are shown in Figure~\ref{fig11} for between 0.95 -– 1.05 d. The effective sampling period peaks shown in  Figure~\ref{fig10} lie neatly between the two spectral leakage peaks for each source shown in Figure~\ref{fig11}. An interesting correlation between the location of the source and the sampling rate can be seen. Sources A, D, and F are all located in the northern hemisphere (i.e. positive declination) and all have an effective sampling period of $\sim$0.979 d. In contrast, sources B, C, and E are all located in the southern hemisphere (negative declination) and all have an effective sampling period of $\sim$0.997 d. This apparent correlation is likely a result of slight differences in the semi-regularity imposed on northern and southern sources, possibly due to factors such as solar X-ray contamination, Earth occultation, and passage of the SAA. As the ASM is either turned off or the data discarded during each of these instances, the process inherently enforces semi-regularity on the sampling rates, possibly differing with source declination.

\begin{table}[hb!]
\begin{center}
\caption{Effective sampling rates $dt$ \& source dec}\label{table1}
\begin{tabular}{lcc}
\hline Source & dec & $dt$ \\
& ($^\circ$) & (d)\\
\hline Crab Pulsar & + 22.01 & 0.9787 $\pm$ 3$\times$10$^{-4}$ \\
\hline LMC X-4 & - 66.37 & 0.9973 $\pm$ 2$\times$10$^{-4}$ \\
\hline Vel X-1 & - 40.60 & 0.9973 $\pm$ 3$\times$10$^{-4}$ \\
\hline Her X-1 & + 35.34 & 0.9785 $\pm$ 5$\times$10$^{-4}$ \\
\hline GX 339-4& - 48.79 & 0.9973 $\pm$ 2$\times$10$^{-4}$ \\
\hline EXO 2030+375 & + 37.64 & 0.9785 $\pm$ 6$\times$10$^{-4}$ \\
\hline
\end{tabular}
\medskip\\
\end{center}
\end{table}

\begin{figure}[htb]
\begin{center}
\includegraphics[scale=0.325, angle=0]{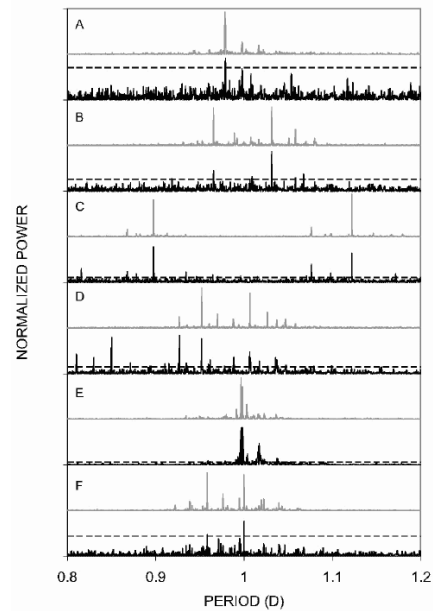}
\caption{Simulated power spectra (grey lines) of the most significant periods present in the light curves of the six test sources with the real power spectra from Figure~\ref{fig3} shown for comparison (black lines). The dashed lines indicate the 99$\%$ significance levels for the real data}\label{fig12}
\end{center}
\end{figure}

The respective input periods were calculated for all observed $\sim$24 h peaks with $>$ 99$\%$ significance using Equations (2) \& (4) and the effective sampling rates. All input periods that were present in the power spectra of Figure~\ref{fig2} were then simulated in light curves with the same sampling as the test sources. For all sources except the Crab pulsar and GX 339-4, these periods correspond either to the reported orbital and super-orbital periods of the sources, or to fractions of these periods. The periods used in the simulations for these sources are as follows: LMC X-4 -- 30 d (super-orbital period); Vela X-1 -- 9.8 d (orbital period), 4.9 d (1/2 orbital period); Her X-1 -- 35 d (super-orbital period), 17 d (1/2 super-orbital period), 1.7 d (orbital period); EXO 2030+375 -- 46 d (orbital period), 23 d (1/2 orbital period), 16 d (1/3 orbital period). As mentioned previously, the cause of the long-term periods in the Crab light curve is unknown, although the two periods used for the simulations (2644 d and 51 d) are possibly related to the solar cycle and precession of the {\it RXTE} orbit. The six periods used for the simulation of GX 339-4 are due to two large scale flaring episodes in the light curve. These state transitions introduce a number of trends and periods into the light curve and result in significant spectral leakage around 24 hr.

The resulting power spectra are shown in Figure~\ref{fig12} for periods between 0.8 -– 1.2 d, with the real power spectra of Figure~\ref{fig3} shown for comparison. Spectral leakage can be seen to account for the most significant peaks around 24 h in each of the test source power spectra. Any periods above the 99$\%$ significance level  that are not present in the simulated power spectra of Figure~\ref{fig12} do not result from the leakage of power from long-term periods present in the real power spectra by the $\sim$1 d effective sampling period. These periods are potentially real, although further analysis is required before a firm conclusion can be drawn. Spectral leakage explains why the values of the $\sim$24 h periods change between sources, and why the spurious period is not present in data from all sources. Each source monitored by the ASM has a unique sampling pattern so that while sources close to each other (as viewed from the Earth) have similar sampling patterns, no two are identical. While it is likely that ASM data for a given source have an enforced semi-regularity and thus an effective sampling rate of $\sim$1 d, it is by no means guaranteed. In order for a significant spurious peak at $\sim$24 h to be present in a light curve two conditions need to be met –-- the sampling rate must have a high enough degree of enforced regularity, and significant real low-frequency modulation(s) must be present. If one or the other of these conditions is not satisfied in a given light curve, the spurious 24 h peaks will not appear.  

\begin{figure}[htb]
\begin{center}
\includegraphics[scale=0.325, angle=0]{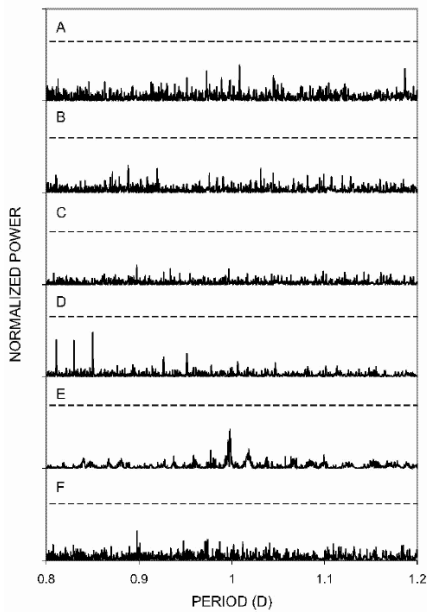}
\caption{Power spectra of the six test source light curves after re-binning using a bin size of 0.3 d. No peaks above the 99$\%$ significance levels (indicated by the dashed lines) appear between 0.8 -- 1.2 d}\label{fig13}
\end{center}
\end{figure}

\section{Conclusions}
In depth analysis of the spurious 24 h periods seen in {\it RXTE} ASM data sets leads us to the conclusion that the observed periods are not dominantly a result of daily variations in background levels. While events such as activation due to SAA passage may be a contributing factor, this is not the primary mechanism. We instead conclude that spectral leakage of low-frequency modulation due to an effective sampling rate of $\sim$1 d is the dominant cause of the spurious signals.

In order to identify a suspect period around 24 h as spurious, a simple technique can be employed. According to Equation (5), re-binning (i.e. changing $1/dt$) the light curve into sample sizes less than 1 d (e.g. 0.3 d averaged observations) should remove all peaks around 24 h due to spectral leakage of any real periods greater than $\sim$0.5 d, leaving behind only real periods (Dieters et al. 2005 employed this technique on GX 339-4 in an attempt to identify the reported $\sim$1 d orbital period). To test this theory we rebinned the light curves of all six test sources using a bin size of 0.3 d. Power spectra of these rebinned light curves are shown in Figure~\ref{fig13}. The significant peaks at $\sim$24 h in each of the power spectra were shifted to 0.3 d, leaving no significant peaks behind. As a further check, we repeated our analysis using a bin size of 0.5 d and found that no significant peaks were present around $\sim$24 h. In addition, any real periods around 24 h should produce spurious peaks of their own which can be identified using Equation (5) and by determining the effective sampling rate through calculating the pseudowindow. These techniques will not distinguish a real period from one possibly produced through daily variations in background levels. However, any spurious periods produced by this effect should have a peak very close to the rotational period of the Earth (i.e. 23.9345 h). Any suspect period not located at this exact value and not a product of spectral leakage could be considered to be real. Furthermore, filtering out all observations influenced by SAA activation should remove any power due to daily variations in background levels, leaving only the real signals and those due to spectral leakage. Additional evidence in the form of periodic variations at other wavelengths or data from other instruments/spacecraft missions would be advised before a firm conclusion is drawn.

\section*{Acknowledgments} 
This research has made use of data obtained through the High Energy Astrophysics Science Archive Research Centre Online Service, provided by the NASA/Goddard Space Flight Centre. We thank Nick Lomb for his patient advice on the use of the Lomb-Scargle analysis technique, Alan Levine for his discussions on the {\it RXTE} ASM instrument, Craig Markwardt for his advice on PCA activation timescales, and Evan Smith for his discussions regarding the {\it RXTE} orbit. We also thank the anonymous referees for their helpful comments.


\end{document}